\documentstyle[preprint,aps]{revtex}

\begin{document}
\draft 
\title
{\bf Band Distributions for Quantum Chaos on the Torus\\}
\author{\bf Itzhack Dana,$^1$ Mario Feingold,$^2$ and Michael Wilkinson$^3$\\}
\address
{$^1$Department of Physics, Bar-Ilan University, Ramat-Gan 52900, Israel}
\address
{$^2$Department of Physics, Ben-Gurion University, Beer-Sheva 84105, Israel}
\address 
{$^3$Department of Physics and Applied Physics, University of Strathclyde, 
Glasgow G4 0NG, U. K.}  
\maketitle

\begin{abstract}
Band distributions (BDs) are introduced describing quantization in a
toral phase space. A BD is the uniform average of an eigenstate
phase-space probability distribution over a band of toral boundary
conditions. A general explicit expression for the Wigner BD is
obtained. It is shown that the Wigner functions for {\em all} of the
band eigenstates can be reproduced from the Wigner BD. Also, BDs are
shown to be closer to classical distributions than eigenstate
distributions. Generalized BDs, associated with sets of adjacent
bands, are used to extend in a natural way the Chern-index
characterization of the classical-quantum correspondence on the torus
to arbitrary rational values of the scaled Planck constant.\newline
\end{abstract}


The classical and quantum dynamics of several nonintegrable model
systems, which have become paradigmatic in the field of Quantum Chaos,
can be reduced to a torus, either in configuration space (e.g., the
Sinai billiard \cite {s,b,hs}) or in phase space (e.g., the ``cat
maps'' \cite{hb,k} and the ``kicked-Harper'' model
\cite{tkh,da,kh,le1,le2,da1,drf}). Quantally, the admissible toral
states have to satisfy proper boundary conditions (BCs), i.e., they
have to be periodic in the torus up to constant Bloch phase
factors. This Letter is concerned with systems for which general Bloch
BCs are defined on a toral phase space: several of such systems have
been discussed in the Quantum-Chaos literature
\cite{hb,k,tkh,da,kh,le1,le2,da1,drf}, although attention has often
been confined to strict periodicity. General Bloch BCs are physically
relevant if the toral phase space represents the unit cell of a
Hamiltonian or mapping which is periodic on phase space: models of
this type occur naturally in problems involving magnetic fields
combined with periodic potentials
\cite{qh,avr,sim,da2,daz,wil,wil1,wil2,a,da3}. A further reason for
studying Bloch BCs is that some physical insight comes from examining
the sensitivity of eigenstates to variation in the BCs
\cite{le1,le2,da1,drf,a}.\newline

It is well known \cite{hb,k,tkh,da,kh,le1,le2,da1,drf} that a
necessary condition for the reduction of phase-space quantum dynamics
to a torus $T_{Q}^{2}$ is that a scaled Planck constant for the
problem, denoted by $\rho =\hbar/2\pi$, assumes rational values: $\rho
=q/p$ ($q$ and $p$ are coprime integers). The ``quantum'' torus
$T_{Q}^{2}$ is $q$ times larger than the torus $T^{2}$ to which the
corresponding classical dynamics can be reduced
\cite{tkh,le2,da1,drf}. For each BC, characterized by the Bloch
wavevector ${\bf w}$, the spectrum consists of precisely $p$ levels
\cite{da1}. As ${\bf w}$ is varied, these levels broaden into $p$
bands labeled by an index $b$. A measure of the sensitivity of the
eigenstates in band $b$ to variations in the BCs is the {\em Chern
index} $\sigma_{b}$ \cite{le1,le2,da1,drf,a}, an integer topological
invariant analogous to the quantum Hall conductance carried by a
magnetic band in a perfect crystal
\cite{qh,avr,sim,da2,daz,wil,wil1,wil2}. The eigenstates may be weakly
dependent on the BCs {\em only} if $\sigma_{b}=0$, a value which may
arise only if $q=1$ \cite{da1}. In this case, where $T_{Q}^{2}=T^{2}$,
one can easily establish a classical-quantum correspondence on the
torus for small $\rho$ \cite {hb,le1,le2}. Several arguments
\cite{le1,le2,a}, supported by numerical evidence, indicate that if
the Husimi distribution of an eigenstate is localized on a classical
regular orbit [e.g., a Kol'mogorov-Arnol'd-Moser torus or a periodic
orbit] the corresponding band has $\sigma_{b}=0$. On the other hand,
eigenstates whose Husimi distribution is spread over the classical
chaotic region should belong to bands with $\sigma_{b}\neq 0$. The
transition from a nearly-integrable regime, where almost all
$\sigma_{b}=0$, to a fully chaotic regime, where almost all
$\sigma_{b}\neq 0$, as a nonintegrability parameter is increased,
takes place via degeneracies between adjacent bands, leading to a
``diffusion'' of the Chern indices \cite{wil3}. The last three
sentences summarize what we call the Chern-index characterization of
the classical-quantum correspondence on the torus.\newline

There is a sense in which eigenstates are not natural for a
characterization of the classical-quantum correspondence, in that they
exhibit rather nonclassical features, due to their association with
the purely quantum quantity ${\bf w}$ and to their generic sensitivity
on this quantity. In particular, the Chern-index characterization
above cannot be extended to the general case $q\neq 1$ on the basis of
the eigenstates \cite{le2,da1}. In order to take into account all the
BCs but, at the same time, to eliminate their individual purely
quantum effects, we propose in this Letter to characterize the
classical-quantum correspondence on a toral phase space by quantities
given by {\em averages} over all the BCs. As a matter of fact, one
such quantity is the Chern index itself, which can be expressed as the
uniform average of the eigenstate two-form with respect to ${\bf w}$
\cite{le1,da1}. Here we introduce the {\em band distribution} (BD),
given by the uniform average of the phase-space probability
distributions (either Wigner or Husimi) for the band eigenstates over
all the BCs. The BD may be viewed as the representative probability
distribution for a level in the torus. We obtain a general explicit
expression for the Wigner BD. A main result of this Letter is
expressed by the formulae (\ref{BWDw}) and (\ref{ABD}) below, by which
the Wigner functions of {\em all} of the band eigenstates are encoded
in a single, analytic function, the Wigner BD. Thus, {\em no
information is lost} about the individual eigenstates by averaging the
Wigner function over the band. Also, it will be demonstrated that the
BDs are ``more classical'' than the distributions of band eigenstates
in several aspects. We also give a generalization of the BD concept,
which is analogous to the smoothing a probability distribution over a
range of energy levels which is used in scar theory \cite{ber}. As a
main application, the generalized BDs are used to extend in a very
natural way the Chern-index characterization of the classical-quantum
correspondence to $q\neq 1$. More details about the Husimi case are
given in Ref. \cite{drf}.\newline

Our phase space is $(u,\ v)$, $[\hat u,\ \hat v]=2\pi {\rm i}\rho$,
and the Hamiltonian or mapping for our system is assumed to be
periodic in this space with a $2\pi \times 2\pi$ unit cell, which is
the classical torus $T^{2}$. If $\rho =q/p$, the quantum torus can be
chosen as $T_{Q}^{2}=[0,\ 2\pi q)\times [0,\ 2\pi )$ and the
$v$-representation of the band eigenstates is \cite{da1}
\begin{equation}
\Psi _{b,{\bf w}}(v)=\sum_{m=0}^{p-1}\phi_{b}(m;\ {\bf
w})\sum_{l=-\infty}^{\infty}\exp [{\rm i}l(w_{1}+2\pi m\rho )/q]\delta
(v-w_{2}+2\pi l/p)\ ,
\label{QES}
\end{equation}
where $b=1,...,\ p$. This state is an eigenfunction of the commuting
phase-space translation operators on $T_{Q}^{2}$, $\hat D_1=\exp({\rm
i}\hat u/\rho )$ and $\hat D_2=\exp({\rm i}p\hat v)$, where $\hat
u=2\pi {\rm i}\rho d/dv$: the eigenvalues are $\exp({\rm i}w_1/\rho)$
and $\exp({\rm i}pw_2)$, implying that ${\bf w}=(w_1,\ w_2)$ is the
Bloch wavevector. Up to phase factors depending only on ${\bf w}$ and
on the Chern index $\sigma_b$, the eigenstates (\ref{QES}) are
periodic in ${\bf w}$ space with a unit cell given by the ``Brillouin
zone'' $BZ=[0,\ 2\pi\rho )\times [0,\ 2\pi /p)$ \cite{da1}. We denote
by $P_{b,{\bf w}}(u,\ v)$ a phase-space probability distribution
(either Wigner or Husimi) for the eigenstates. We define the {\em band
distribution} (BD) for band $b$ by
\begin{equation}
P_{b}(u,\ v)=\frac{1}{|BZ|}\int_{BZ}d{\bf w}\, P_{b,{\bf w}}(u,\ v)\ ,
\label{BD}
\end{equation}
where $|BZ|=4\pi ^{2}q/p^{2}$ is the area of the Brillouin zone. A
more explicit expression for $P_{b}(u,\ v)$ can be obtained in the
Wigner case. We find in this case, using (\ref{QES}),
\begin{eqnarray}
P_{b,{\bf w}}(u,\ v) &\equiv &\frac{1}{2\pi ^{2}\rho }\int_{-\infty
}^{\infty}dv^{\prime}\exp ({\rm i}uv^{\prime}/\pi\rho ) \Psi_{b,{\bf
w}}^{\ast}(v-v^{\prime}) \Psi_{b,{\bf w}}(v+v^{\prime}) \nonumber \\
&&=\sum_{r=-\infty}^{\infty}\sum_{s=-\infty}^{\infty}A_{b}(r,\ s;\
{\bf w})\ \delta (u-w_{1}-r\pi\rho )\ \delta (v-w_{2}-s\pi /p)\ ,
\label{BWDw}
\end{eqnarray}
where the first equality defines the Wigner function, and where
\begin{equation}
A_{b}(r,\ s;\ {\bf w})=\frac{p}{4\pi}\sum_{m=0}^{p-1}\exp \left[
\frac{{\rm i}\pi s(r-2m)}{p}\right] \phi_{b}(m;\ {\bf w})
\phi_{b}^{\ast}[(r-m)\ \mbox{mod }p;\ {\bf w}]\ .
\label{Ab}
\end{equation}
The result (\ref{BWDw}) shows that the support of the Wigner function
for an eigenstate is, quite generally, a lattice in phase space, which
is shifted uniformly by shifting ${\bf w}$. This generalizes the
result of Hannay and Berry \cite{hb} for ${\bf w=0}$ (strict
periodicity) to arbitrary BCs. Using (\ref{BWDw}) in (\ref{BD}), we
obtain the following expression for the Wigner BD:
\begin{equation}
P_{b}(u,\ v)=\frac{p^{2}}{4\pi ^{2}q}\sum_{r=0,1}\sum_{s=0,1}A_{b}(r,\
s;\ w_{1}=u-r\pi\rho ,\ w_{2}=v-s\pi /p)\ . \label{BWD}
\end{equation}

Equation (\ref{BWD}) will now be {\em inverted} to express $A_{b}(r,\
s;\ {\bf w})$ in terms of the Wigner BD. To this end, one notices
first that, due to the strict periodicity of Eq. (\ref{BWDw}) in the
Brillouin zone, the right-hand side of (\ref{BWD}) will not change if
the summations are performed over $r={\bar r},\ {\bar r}+1$ and
$s={\bar s},\ {\bar s}+1$, where ${\bar r}$ and ${\bar s}$ are
arbitrary integers. Next, one writes Eq. (\ref{BWD}) with $(u,\ v)$
replaced by $(u+kq\pi ,\ v+l\pi )$ for $k=0,\ 1$ and $l=0,\ 1$,
choosing ${\bar r}=kp$, ${\bar s}=lp$ in the four cases. Using
(\ref{Ab}), one then finds that $P_b(u+kq\pi ,\ v+l\pi )$ is given by
the right-hand side of Eq. (\ref{BWD}) with the extra factor
$(-1)^{ks+lr+klp}$ under the summation signs. The resulting four
equations can be easily solved for $A_{b}(r,\ s;\ {\bf w})$:
\begin{equation}
A_{b}(r,\ s;\ {\bf w})=\frac{\pi^{2}q}{p^{2}}
\sum_{k=0,1}\sum_{l=0,1}(-1)^{ks+lr+klp}P_b(u+kq\pi ,\ v+l\pi )\ ,
\label{ABD}
\end{equation}
where $u=w_1+r\pi\rho$ and $v=w_2+s\pi /p$. Together with the latter
expressions, Eq. (\ref{ABD}) shows that the Wigner functions
(\ref{BWDw}) of all the band eigenstates are simply encoded in the
single, smooth phase-space function $P_b(u,\ v)$. Thus, no information
about (\ref{BWDw}) is lost by performing the average in (\ref{BD}), as
this information is fully recoverable from the Wigner BD.\newline

We now discuss properties of the BD (\ref{BD}), including aspects in
which it is ``more classical'' than $P_{b,{\bf w}}(u,\ v)$. First, we
remark that $P_b(u,\ v)$, in particular (\ref{BWD}), is a smooth
function, unlike the Wigner function (\ref{BWDw}) of the
eigenstates. Also, using the relation $e^{{\rm i}r({\hat
v}-w_{2})/\rho }\Psi_{b,{\bf w}}(v)=\Psi_{b,w_{1}-2\pi r,w_{2}}(v)$
($r$ integer) \cite{da1} in (\ref{BD}), and noticing that $\exp ({\rm
i}r{\hat v}/\rho )$ is just a translation of $u$ by $2\pi r$, we
easily find that $P_{b}(u,\ v)$ is periodic with unit cell $T^{2}$
(the classical torus) for {\em general} $q$. This is in contrast with
$P_{b,{\bf w}}(u,\ v)$, whose unit cell of periodicity is the quantum
torus $T_{Q}^{2}$. We can then impose on $P_{b}(u,\ v)$ the
normalization condition $\int_{T^{2}}du\, dvP_{b}(u,\ v)=1$, making
$P_{b}(u,\ v)$ analogous to a classical probability distribution in
the phase space $T^{2}$. Using the periodicity of $P_{b}(u,\ v)$ in
Eq. (\ref{ABD}), it is easy to see that at any four points $(u+kq\pi
,\ v+l\pi )$ ($k=0,\ 1$, $l=0,\ 1$) in $T_{Q}^{2}$, $P_{b,{\bf w}}(u,\
v)$ in (\ref{BWDw}) assumes values differing at most in sign, a fact
which was observed in Ref. \cite{hb} in the particular case of ${\bf
w=0}$ (and $q=1$). This means that only a quarter of the values
assumed by (\ref{BWDw}) in $T_{Q}^{2}$ may be independent. This rather
nonclassical property of (\ref{BWDw}) is generally not possessed by
the Wigner BD.\newline
   
Next, consider $P_{b,{\bf w}}(u,\ v)$ in the Husimi case: $P_{b,{\bf
w}}(u,\ v)=|\Psi_{b,{\bf w}}(u,\ v)|^{2}$, where $\Psi_{b,{\bf w}}(u,\
v)$ is the coherent-state representation of $|\Psi_{b,{\bf
w}}\rangle$. For given $(b,\ {\bf w)}$, $P_{b,{\bf w}}(u,\ v)$ always
assumes $p$ zeros $(u_{0,j},\ v_{0,j})$ ($j=1,...,\ p$) in $T_{Q}^{2}$
\cite{le1,da1,drf}. These zeros make $P_{b,{\bf w}}(u,\ v)$ rather
``nonclassical'', for example, they do not allow $P_{b,{\bf w}}(u,\
v)$ to approach, in the semiclassical limit, the microcanonical
uniform distribution in a strong-chaos regime \cite{le1,le3}. On the
other hand, the Husimi BD {\em never vanishes} [$P_{b}(u,\ v)>0$ in
$T^{2}$], simply because the $p$ zeros $(u_{0,j},\ v_{0,j})$ generally
vary with ${\bf w}$ and the definition (\ref{BD}) involves an
integration over all ${\bf w}$.\newline

Finally, we discuss the important case of bands with Chern index
$\sigma_{b}=0$, which is possible only when $q=1$ \cite{da1}. In this
case, $|\Psi_{b, {\bf w}}\rangle$ can be written as a symmetry-adapted
sum \cite{da2,wil},
\begin{equation}
|\Psi _{b,{\bf w}}\rangle =\sum_{l_{1},l_{2}=-\infty}^{\infty}
e^{-{\rm i}p(l_{1}w_{1}+l_{2}w_{2})} \hat D_{1}^{l_{1}}\hat
D_{2}^{l_{2}} |\varphi_{b}\rangle \ , \label{sas}
\end{equation}
where $|\varphi_{b}\rangle$ is some square-integrable state, which is
analogous to a Wannier function \cite{da2,wil,wil2}. Inserting
(\ref{sas}) into (\ref{BD}), we easily obtain a general exact
expression for the BD:
\begin{equation}
P_b(u,\ v)=\sum_{l_1,l_2=-\infty}^{\infty} P_{\varphi_b}(u+2\pi l_1,\
v+2\pi l_2)\ ,
\label{BD0}
\end{equation}
where $P_{\varphi_b}(u,\ v)$ is the Wigner or Husimi function of the
Wannier state $\vert\varphi_b\rangle$. While the Wannier function is
not invariant under gauge transformations in which the Bloch states
are multiplied by $\exp [{\rm i}\theta (w_1,\ w_2)]$ \cite{wil2}, the
BD (\ref{BD0}) is gauge invariant. In a nearly-integrable situation
and in a semiclassical regime, the Husimi $P_{\varphi_b}(u,\ v)$ is
well localized on a classical regular orbit, provided band $b$ is well
separated from neighboring bands [$P_{\varphi_b}(u,\ v)$ is then the
``quasi-mode'' of Ref. \cite{fau}]. In the semiclassical limit
$\rho\rightarrow 0$, $P_{\varphi_b}(u,\ v)$ tends point-wise to zero
outside the classical orbit \cite{fau}. Similarly, the BD (\ref{BD0})
in the Husimi case tends point-wise to zero outside the periodic
repetition of the classical orbit on all unit cells $(l_{1},\
l_{2})$. It is therefore a periodic version of the quasi-mode Husimi
density, appropriate for a toral phase space. The difference
$P_{b,{\bf w}}(u,\ v)-P_{b}(u,\ v)$ is the sum of the overlaps of the
quasi-mode with the translated quasi-mode in all unit cells $(l_{1},\
l_{2})\neq (0,\ 0)$, and it is of a purely quantum nature. This
clarifies the classical nature of the BD in this case.\newline

In some important cases, it is necessary to generalize the BD concept
by averaging over more than one band, usually over a set of $N$
adjacent bands $b=b_{1},...,\ b_{N}$. This set may be considered as a
single entity, a {\em generalized band} (GB), which can be
characterized by its total Chern index, $\sigma_{\rm GB}\equiv
\sum_{b=b_{1}}^{b_{N}}\sigma_{b}$, and by the {\em generalized} BD
associated with it, $P_{\rm GB}(u,\
v)=N^{-1}\sum_{b=b_{1}}^{b_{N}}P_{b}(u,\ v)$. The further averaging
over bands should give a ``more classical'' BD, as when smoothing over
many levels in a general quantum system \cite{ber}. The ``maximal''
smoothing is, of course, that over all the $p$ bands. From the
completeness of the eigenstates (\ref{QES}), we find in this case that
$p^{-1} \sum_{b=1}^{p}P_{b}(u,\ v)=(4\pi ^{2})^{-1}.$ Thus, as one
could expect, the generalized BD in this case is just the uniform
distribution in phase space.\newline

The use of generalized BDs is quite natural, for example, near a
degeneracy between a pair of bands. In fact, precisely at the
degeneracy point it is usually not useful to consider the two bands
separately, and they must be treated as one single entity (the GB). It
is well known \cite{sim} that the Chern indices of the two bands
generically vary by $\pm 1$ (for $q=1$) across the degeneracy, leaving
their total Chern index unchanged. Similarly, one can show \cite{drf}
that the generalized BD for the two bands is approximately conserved
across the degeneracy, despite the fact that the separate BDs may vary
drastically.\newline

We are now ready to present a main application of the BD concept. We
show how generalized BDs can be used to extend in a natural way the
Chern-index characterization of the classical-quantum correspondence
on the torus \cite {le1,le2,da1} to general rational values $q'/p'$ of
$\rho $ near the special values of the form $1/p$ for which this
characterization was originally formulated. Our basic assumption is
that the renormalization-group approach developed in \cite{wil1,wil2},
which was applied to the investigation of the spectrum of a general
class of Hamiltonians on the torus, is applicable to the band spectrum
of our nonintegrable system. This assumption has been verified
numerically for the kicked Harper model on a broad interval of the
nonintegrability parameter \cite{drf,next}. Let $\rho ^{\prime}=
q^{\prime}/p^{\prime}$ be a rational number sufficiently close to
$\rho =q/p$ and such that $p^{\prime}\gg p$. From Ref. \cite{wil1}, we
know that the $p^{\prime}$ bands for $\rho ^{\prime }$ can be grouped
into $p$ ``clusters'' of adjacent bands, where each cluster $C_{b} $
is associated in a natural way with a band $b$ for $\rho
=q/p$. Namely, the energy or quasienergy interval covered by the bands
in $C_{b}$ is relatively close to that covered by band $b$ and the
total Chern index $\sigma (C_{b})$ of $C_{b}$ is equal to
$\sigma_{b}$. The spectrum and eigenstates in $C_{b}$ can be
calculated approximately from an effective Hamiltonian $H_{\rm eff}$,
obtained by properly quantizing the band function for band
$b$.\newline

The existence of $H_{\rm eff}$ means that the space of states in band
$b$ approximately coincides with the space of states in $C_{b}$. In
other words, the projection operator for band $b$ is approximately
equal to that for $C_{b}$:
\begin{equation}
\frac{1}{|BZ|}\int_{BZ}d{\bf w}|\Psi_{b,{\bf w}}\rangle\langle\Psi
_{b,{\bf w}}|\approx \frac{1}{N_{b}}\sum_{b^{\prime}=
d(b)}^{d(b)+N_{b}-1}\frac{1}{|BZ^{\prime}|}\int_{BZ^{\prime}} d{\bf
w^{\prime}}|\Psi_{b^{\prime},{\bf w^{\prime}}}^{\prime}
\rangle\langle\Psi_{b^{\prime},{\bf w^{\prime}}}^{\prime}|\ ,
\label{po}
\end{equation}
where all the primed quantities refer to $\rho ^{\prime}$, $N_{b}$ is
the number of bands in $C_{b}$, and $d(b)$ is the label of the lowest
band in $C_{b}$. We immediately obtain from (\ref{po}) that
\begin{equation}
\frac{1}{|BZ|}\int_{BZ}d{\bf w}|\Psi_{b,{\bf w}}(u,\ v)|^{2}\approx
\frac{1}{N_{b}}\sum_{b^{\prime}=
d(b)}^{d(b)+N_{b}-1}\frac{1}{|BZ^{\prime}|}\int_{BZ^{\prime}} d{\bf
w^{\prime}}|\Psi_{b^{\prime},{\bf w^{\prime}}}^{\prime} (u^{\prime},\
v^{\prime})|^{2}\ , \label{BDc}
\end{equation}
where the variables $(u^{\prime},\ v^{\prime})$ for $\rho ^{\prime}$
are related to the variables $(u,\ v)$ for $\rho$ by $(u^{\prime},\
v^{\prime})=\sqrt{\rho /\rho ^{\prime}}(u,\ v)\,$. Eq. (\ref{BDc})
shows that the Husimi BD for band $b$ is approximately equal to the
generalized Husimi BD $P_{C_{b}}^{\prime} (u^{\prime},\ v^{\prime})$
for cluster $C_{b}$. An analogous approximate equality in the Wigner
case can be similarly established. In the limit $\rho
^{\prime}\rightarrow \rho $, the space of the cluster becomes
identical to that of band $b$ [the approximate equality in (\ref{po})
is replaced by an equality], and $P_{C_{b}}^{\prime}(u^{\prime},\
v^{\prime})\rightarrow P_{b}(u,\ v)$.\newline

The most important reference values of $\rho$ are those with $q=1$,
for which the Chern-index characterization of the classical-quantum
correspondence is well established \cite{le1,le2,da1}. For these $\rho
$'s, Eq. (\ref{BDc}) implies that if the BD for band $b$ is
concentrated, in a semiclassical regime, on a regular classical orbit
[$\sigma_{b}=\sigma (C_{b})=0$] or on the classical chaotic region
[$\sigma_{b}=\sigma (C_{b})\neq 0$], the same will be true for the
generalized BD for $C_{b}$. One can also show that general cluster
states, characterized by a well-defined value of ${\bf w}$, are
``weakly'' or ``strongly'' sensitive to variations in ${\bf w}$
depending on whether $\sigma (C_{b})=0$ or $\sigma (C_{b})\neq 0$
\cite{drf}. The Chern-index characterization of the classical-quantum
correspondence on the torus is thus extended to $\rho ^{\prime}$
sufficiently close to $\rho =1/p$ by replacing single bands $b$ with
the corresponding clusters $C_{b}$.\newline

In conclusion, the BD concept introduced in this Letter was shown to
exhibit interesting and useful properties: (a) The Wigner functions of
the band eigenstates are fully recoverable from the Wigner BD. (b) At
the same time, the BDs are ``more classical'' than eigenstate
distributions in several aspects. (c) Generalized BDs allow to extend
in a natural way the Chern-index characterization of the
classical-quantum correspondence to $q\neq 1$.\newline

We remark that the BD concept could be further developed in several
directions. Smoothing a quantum probability distribution over a range
of energy levels is important in the theoretical study of scars using
the semiclassical periodic-orbit theory \cite{ber}. In the case of the
BD, the smoothing is performed over the continuous range of one band,
corresponding essentially to a {\em single} level in the framework of
a toral phase space. This ``minimal'' smoothing is performed just for
the sake to eliminate the purely quantum effects of individual
BCs. Using the adaptation of periodic-orbit theory to the framework of
a toral phase space \cite{lem}, it may be possible to achieve a better
understanding of the nature of BDs and generalized BDs in the
semiclassical limit. Also, a more complete characterization of the
classical-quantum correspondence on the phase-space torus should be
achieved by introducing, in addition to the Chern index and the BDs,
new quantities which are also naturally defined as averages over all
the BCs. It should also be interesting to determine whether the
distributions for the band eigenstates are also recoverable from the
Husimi BD, and to extend the representation (\ref{BD0}) of the BD in
terms of Wannier states to the case $\sigma_b\neq 0$, using results
from Ref. \cite{wil2}.\newline

{\bf Acknowledgments\\}

We would like to thank J. Zak, P. Leboeuf, A. Voros, M. V. Berry, and
Y. Rutman for comments and discussions. ID and MF acknowledge support
from the Israel Science Foundation. MW acknowledges support from the
EPSRC, grant GR/L/02302, and the hospitality of Ben-Gurion and
Bar-Ilan Universities.

\end{document}